# Interfacial Atomic Number Contrast in Thick TEM Samples

Aniruddha Dutta * & Helge Heinrich †

*Department of Physics, University of Central Florida, Orlando, Florida, U.S.A.
†Department of Materials Science and Engineering, University of Virginia, 395 McCormick Road, Charlottesville, VA 22904, U.S.A.

## Abstract

The atomic number contrast imaging technique reveals an increase in intensity at interfaces of a high and low-density material in case of relatively thick samples. Elastic scattering factors and absorption coefficients are incorporated in a probabilistic model to study atomic contrast occurring at the interface of two materials when the High-Angle Annular Dark-Field (HAADF) detector is used in the Scanning TEM (STEM) mode. Simulations of thick samples reveal that electrons traverse from a higher density material to a lower density material near the interface which increases the HAADF-STEM signal. This effect is more dominant in TEM samples of thickness greater than 100 nm and the increase in signal occurs up to 20 nm from the interface. The behavior of electrons near the interface is explained by comparing the simulation results with experimental TEM micrographs in the HAADF-STEM mode.

## 1. Introduction

While imaging cross-sections of multilayered samples prepared by the Focused Ion Beam (FIB) technique, we observed in some cases an increase of the HAADF intensity at interfaces when using the Scanning Transmission Electron Microscopy (STEM) mode. This effect was only observed in FIB samples with thicknesses of 100 nm or more. For imaging of samples in the STEM mode an increase in High-Angle Annular Dark-Field (HAADF) intensity usually can be related to materials with higher atomic number and/or density, but no such layers were found using Energy-Dispersive X-ray Spectroscopy (EDXS) or Electron Energy-Loss Spectroscopy (EELS) techniques. An example of a bright HAADF-STEM contrast at the interface between a gold and a titanium layer as well as between a titanium and a $LiTaO_3$ layer is shown in Figure 1. Thick STEM samples reveal an increase in intensity at the interfaces of high and low-density materials when using an HAADF-STEM detector.

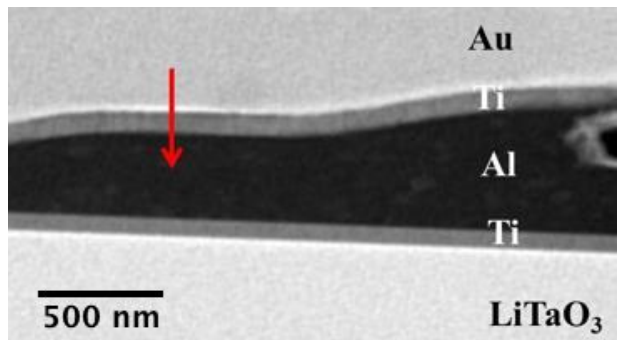

Fig. 1. HAADF-STEM micrograph of a multilayered sample provided by Qorvo Inc.. The red arrow indicates the profile scan direction from Au to Ti Layer shown in Fig. 2.

Fig. 1. shows a HAADF-STEM micrograph of a FIB cross section from a multilayered sample provided by Qorvo Inc. in Apopka, Florida. The profile scan in the STEM mode reveals an increase in intensity (marked by the red circle in Fig. 2.) just near the Au and Ti interface. The micrograph suggests that there is a higher density material between Au and Ti as the HAADF-STEM signal increases with increasing atomic number. Au being a higher density material than Ti, the above effect has however to be explained by the specifics of scattering of electrons near the interface, as EDXS measurements do not reveal any other elements besides Au and Ti. A local higher density of Au can also be excluded, as Au does not form any compound with Ti that has a significantly

higher atomic density. This paper quantitatively describes the scattering effects near the interface of a high and low-density material using simulations of electrons scattering. These simulations are compared with intensity signals from experimental micrographs.

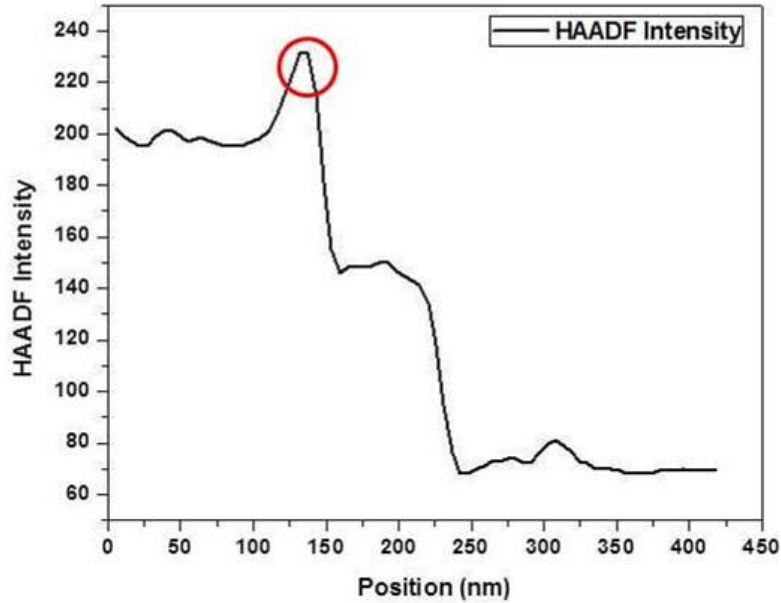

Fig. 2. Profile scan showing an increase of the HAADF signal near the interface of Au and Ti. Left: Au, middle: Ti, right: Al.

Incoherent scattering plays an important part in the formation of HAADF-STEM images. Imaging with STEM is capable of measuring sizes/masses/thicknesses of materials at the atomic scale[1-4] and allows for tomography applications[5,6]. In HAADF-STEM a finely focused beam is scanned across the sample while the HAADF detector collects only electrons scattered to high angles. Due to incoherent scattering, thermal diffuse scattering (TDS) is an important part in HAADF images. With STEM contrast simulations TDS can be accounted for in the HAADF-STEM micrographs. Researchers[7-11] have studied TDS with different approaches through simulations. Quantification of image intensities per pixel from STEM micrographs[12] have revealed information about inelastic scattering and the percentage of the incident beam intensity being absorbed. The multi-slice simulation method[13] considers both dynamically and thermally scattered electrons. For these simulations, a model crystal is sectioned into slices and the wave function of the electron beam is modified by the atomic potential and the angle-dependent propagator function in each slice. The scattered intensities obtained in the diffraction plane are integrated over the detector area for information on each STEM probe position in real space.

A single or dual beam Focused Ion Beam (FIB) can be used to prepare cross-sectional samples for TEM with typically a few 10 nm of thickness. Some parts of these FIB cuts may still be thicker than 100 nm though, especially with ex-situ lift-out samples. This paper addresses imaging artefacts at interfaces occurring for sample thicknesses over 100 nm and provides a new approach to simulate these artefacts.

Multi-slice simulations of the HAADF-STEM contrast however are not useful for thick samples: For a 500 nm thick specimen, we would need frame sizes of the order $2_{16}$ x $2_{16}$ pixels for which Fourier forward and backward transformations have to be performed repeatedly throughout the multi-slice process. The reason for this extreme number of pixels is twofold: With the HAADF-STEM detector, electrons scattered to high angles are contributing to the signal. Angles as high as 250 mrad need to be included in simulations. This sets the minimum size of the real-space image to $2 * 0.250 * 500\ nm = 250\ nm$. In Fourier space (diffraction space) this translates to a pixel size of $\frac{1}{(250 nm)}$. On the other hand, to include 300 kV electrons scattered to 250 mrad in any direction for multi-slice simulations, the size of the Fourier image frame has to be $2 * 2 * \frac{\sin(0.125)}{wavelength} = 250\ nm^{-1}$. This extreme requirement of computing power renders the multi-slice method ill-suited for very thick STEM samples considering the fact that the frame size needs to be increased proportional to the sample thickness.

In this work, we apply an approach commonly used in simulations of electron-solid interactions for Scanning Electron Microscopy (SEM). The above-mentioned issues of extreme frame sizes can thus be avoided. Atomic densities of materials are used while the crystal structure is neglected. For each electron, the probability of electron scattering has a radial symmetry for each electron position and for each beam direction. The probability is derived from scattering cross sections of each element. For contrast simulations of thick samples this model gives a good estimate of scattering and absorption for each slice considering thermal diffuse scattering (TDS).

**2. Sample Preparation**

Qorvo Inc., Florida provided us with multilayered samples from which cross-sections were prepared for TEM. We used a single-beam FEI FIB 200 to prepare conventional cross sections for TEM as well as wedge-shaped cross-sectional samples. The wedge-shaped samples were prepared to study the effect of thickness changes on the HAADF-STEM contrast of interfaces. A typical wedge-shaped sample is (10-12) μm in length and 4 μm in width on the thicker end. A top-view FIB micrograph of a wedge-shaped sample is shown in Fig. 3.

A TECNAI F30 TEM, equipped with a field emission source operating at 300 kV with a point to point resolution of 0.2 nm was used to study the samples. A Fischione HAADF detector with a contrast/brightness setting of 12.5% and 46.875% and a camera length of 80 mm was consistently used[14-16] in the STEM mode. This camera length yields a HAADF detector range for the total scattering angle of (80-460) mrad. These particular settings allowed for a quantitative calibration of the contrast of the specimen. The wedge-shaped samples were analyzed in the HAADF-STEM mode with the detector intensity measured as a function of thickness ranging from 0 nm to several microns. We then compared the measured detector intensities with probabilistic model results for scattering of electrons as described in section 3.1 for quantitative analysis.

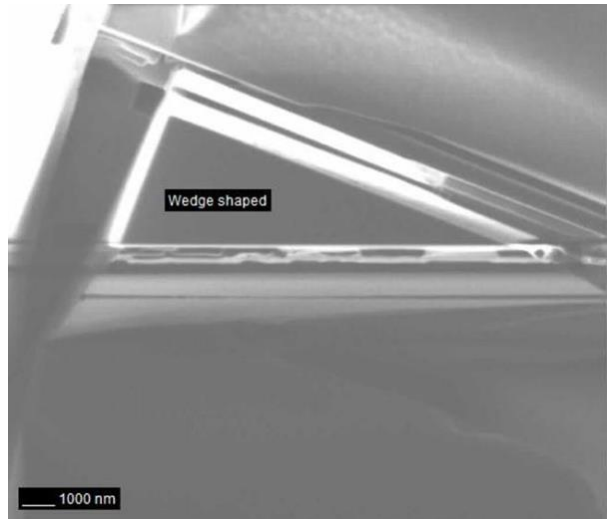

Fig. 3. Top view of a wedge-shaped sample using Ga ion imaging in the FIB after initial cutting of the sides.

## 3. Mathematical Model

An incident electron can be scattered elastically or inelastically by the atoms of the solid. Yoshioka (1978) has shown that the interaction of the incident electron with the solid in the form of a complex optical potential can explain the effects of inelastic scattering. During an inelastic scattering event, the incident electron loses a part of its energy and is removed from the elastic signal. The probability of an electron regaining its lost energy and reappearing is very small[17] and thus inelastic scattering events contribute to an imaginary part of the optical potential. The complex potential for the interaction of electrons with the solid may be represented[18] by the following equation:

$$V(r) = \sum_{i=1}^{N} \varphi_i(r - r_i) \quad (1)$$

where, N represents the number of atoms of the solid, $\varphi_i(r)$ is the contribution from the $i$th atom and is related to the atomic scattering factor $f^{(e)}(s)$ by the following Fourier transform:

$$\varphi(\boldsymbol{r}) = \left(\frac{h^2}{4m_0\pi^3}\right) \int f^{(e)}(\boldsymbol{q}) \exp(2i\boldsymbol{q}.\boldsymbol{r}) \, d^3\boldsymbol{q} \qquad (2)$$

Here, $m_0$ is the rest mass of an electron and h is the Planck constant. Some of the inelastically scattered electrons lose a significant fraction (several keV) of their initial energy. The chromatic aberration of the post-sample magnetic lenses inside the TEM which form the diffraction pattern in the plane of the HAADF detector as well as multiple scattering events deflects these electrons over a large angular range and are less likely than elastically scattered electrons to appear in the detector channel of the microscope.

Relativistic Hartree-Fock wave functions have been used by several authors[19, 20] to calculate the real part of the atomic potential. Doyle & Turner[20] provided the following analytical expression for the elastic scattering factor:

$$f^e(s) = \sum_{i=1}^{n} a_i \exp(-b_i s^2) + c \qquad (3)$$

where $a_i$, $b_i$ and c are parameters determined by curve fitting procedures[21]. In equation (3), $s = \frac{\sin\theta}{\lambda}$, where $\theta$ is half the angle of scattering ($2\theta$ total scattering angle) and $\lambda$ is the wavelength of the incident electrons.

Several research groups have computed the elastic scattering factors[20,22-23]. We use elastic scattering factors of elements computed by *Peng et al.*[18] based on the combined modified simulated-annealing procedure[24-25]. In the work of *Peng et al.*[18] computed values for the range $s = 0 \text{ to } 6 \text{ Å}^{-1}$ are described. For this project, we have used these coefficients for elastic scattering factors. *Peng et al.*[18] deals only with those electrons which get removed from the coherent scattering channels. However, these electrons, to a large part, are still detected in the current

HAADF-STEM experiments as they constitute the diffuse background between Bragg reflections in the diffraction plane. Thus, regarding absorption, the Debye model used by *Peng et al.*[26] is not well suited for the current set of experiments. A simpler approach can be used to determine the absorptive scattering factors of different elements based on the following principles: Previous research[27] shows that the HAADF-STEM intensity increases linearly with thickness for thin samples, which is given by $I = KFt$ where F is the fraction of electrons scattered per nanometer of sample thickness (*t*). For higher sample thicknesses the electrons undergo multiple scattering events and the normalized HAADF-STEM intensity would reach a saturation value, if we had no absorption which can be represented in the form of $I/I_0 = \frac{K}{I_0}[1 - \exp(-Ft)]$. Here $K$ is a prefactor and $K/I_0$ represents the fraction of electrons passing through the sample and reaching the HAADF detector plane and $I_0$ is the initial beam intensity. We use a single absorption parameter (µ) (specific to a material) which accounts for the electrons lost to any detection in thick samples due to multiple scattering to even higher angles than the angular range of the HAADF detector including backscattering. The final thickness-dependent HAADF-STEM intensity can be written as:

$$\frac{I}{I_0} = \frac{K}{I_0}[(1 - \exp(-Ft))\exp(-\mu t)] \qquad (4)$$

The absorption parameter µ is determined from experimental data of wedge-shaped samples as described in section 4.

*3.1. Probabilistic Model for Scattering of Individual Electrons*

A Python 3.5 programming code was developed to simulate the HAADF-STEM signal as a function of sample thickness. Individual electrons from the incident electron beam interact with the electrons and nuclei of the material thereby undergoing multiple scattering events. This probabilistic scattering model is applied for a high number of individual electrons for contrast simulations. A convergent electron beam is scanned across an interface, where several thousand up to a million individual electrons are transmitted through the sample for each beam position in the simulation. For each thin slice of the sample (1-10 nm), the individual electrons get scattered with probabilities that are derived from the elastic and absorptive coefficients of the two materials (e.g., W and $SiO_2$). *Peng et al.*[18] parameters (elastic coefficients) have been used in this program, as well as absorptive parameter ($\mu$) to determine scattering cross-sections of both materials.

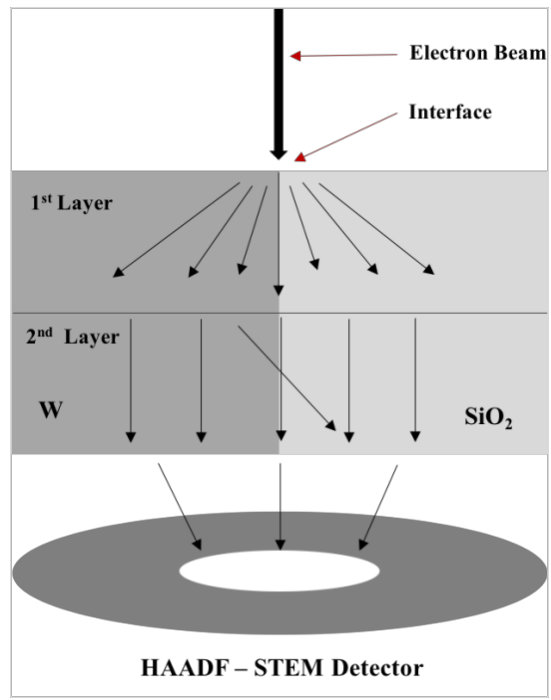

Fig. 4. Schematic of the electron beam entering a sample at the interface between two materials of higher and lower density.

The HAADF-STEM signal is determined for several scan positions at different sample thicknesses in both materials and at different distances from the interface. Fig. 4. shows the schematic of the electron path near an interface. The program employs these absorption and scattering probabilities at every thickness step. The path of each electron is tracked as the electron passes through the sample undergoing multiple scattering events. The tracking is done both in real space as well as in momentum space. The real space position of the electron is recorded to track the position of the electron near the interface and to determine if it is in the low or high density material. This is important as different scattering probabilities are used for both materials. The program also tracks the angle at which the electron traverses the materials. This corresponds to the x-y components of the electrons momentum vector and to the scattering angle of electrons in the diffraction plane.

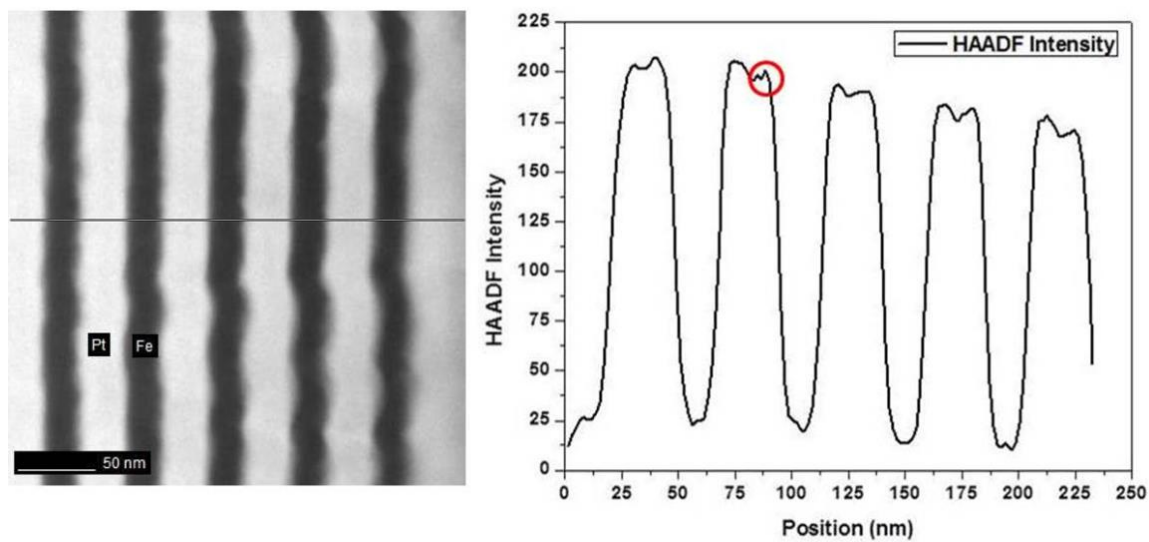

Fig. 5. (Left) Edge-on HAADF STEM micrograph of a multilayer system six Pt (28 nm) and six Fe (22 nm) layers provided by Dr. Bo Yao and Dr. Kevin Coffey (AMPAC, University of Central Florida). (Right) Line scan across the Pt & Fe layers showing an increase in the intensity at the interface.

Fig.5. illustrates another example showing the increase of atomic contrast near the interface of Pt and Fe layers. Fig. 5. (Left) shows the HAADF-STEM micrograph of six Pt (28 nm) and six Fe (22 nm) layers. A profile scan across the layers confirms an increase in the intensity near the interface (red circle) of Pt and Fe layers as shown in Fig. 6. (Right).

## 4. Simulations

A wedge-shaped sample of Fe was prepared to determine one of these element specific absorption parameters. The HAADF-STEM signal was measured experimentally and compared to the HAADF signal simulation as described in section 3.

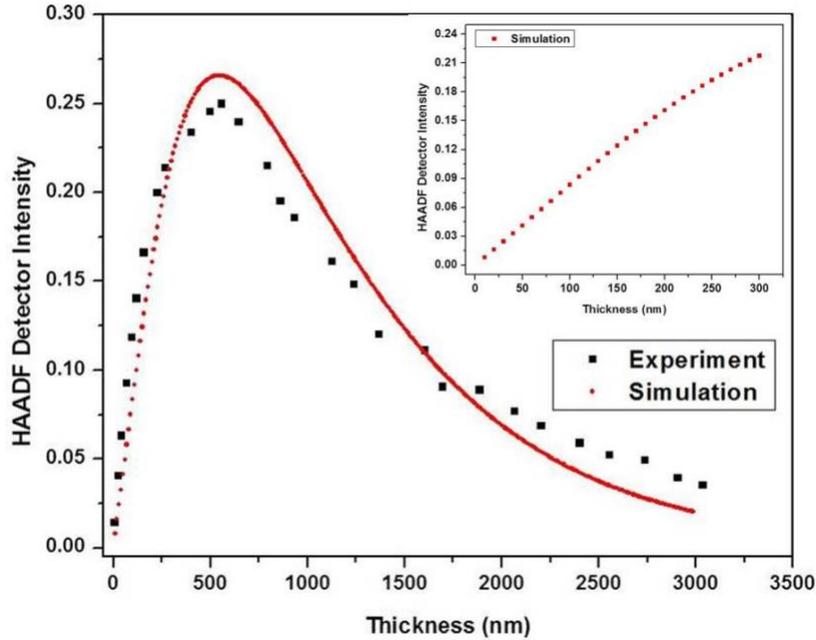

Fig. 6. HAADF-STEM intensity vs thickness of a wedge-shaped Fe sample for both experiment and simulation using the modified multislice method. The inset shows the initial slope from the simulated curve.

Fig. 6. shows the comparison of measured experimental values and simulated values using equation 4 of the HAADF-STEM intensity for an Fe sample of increasing thickness. The inset in Fig. 6. shows the initial slope of the HAADF signal for simulated values.

The absorption parameter $\mu = (1.0 \pm 0.2) * 10^{-4}$ nm-1 is calculated from the best fit of multiple simulations to the experiment. The initial slope of the simulated curve in Fig. 6. is $\varepsilon = (1.12 \pm 0.08) * 10^{-3}$ nm-1 and $F = (8.4 \pm 1.1) * 10^{-4}$ nm-1. Fig. 7. shows the HAADF-STEM intensity vs. thickness of Cu sample obtained from experiments and simulations. The HAADF signal increases with thickness and reaches a maximum around 850 nm. The absorption parameter in Cu reduces the HAADF signal up to the maximum sample thickness of 4 μm. The absorption parameter determined from fitting the experimental data yields $\mu = (4.6 \pm 0.04) * 10^{-4}$ nm-1. The initial slope ε calculated from simulations is $(1.11\pm 0.11) *10^{-3}$nm-1 and $F = (1.53 \pm 0.02) * 10^{-3}$nm-1. The inset in Fig. 7. shows that the simulated curve reproduces the experimental curve well. Small deviations arise from surface roughness and ion implantation of the wedge-shaped sample due to back deposition of gallium from FIB processing. The simulations and experiment also confirm the qualitative model described in equation 4 for the thickness dependence of the HAADF-STEM intensity.

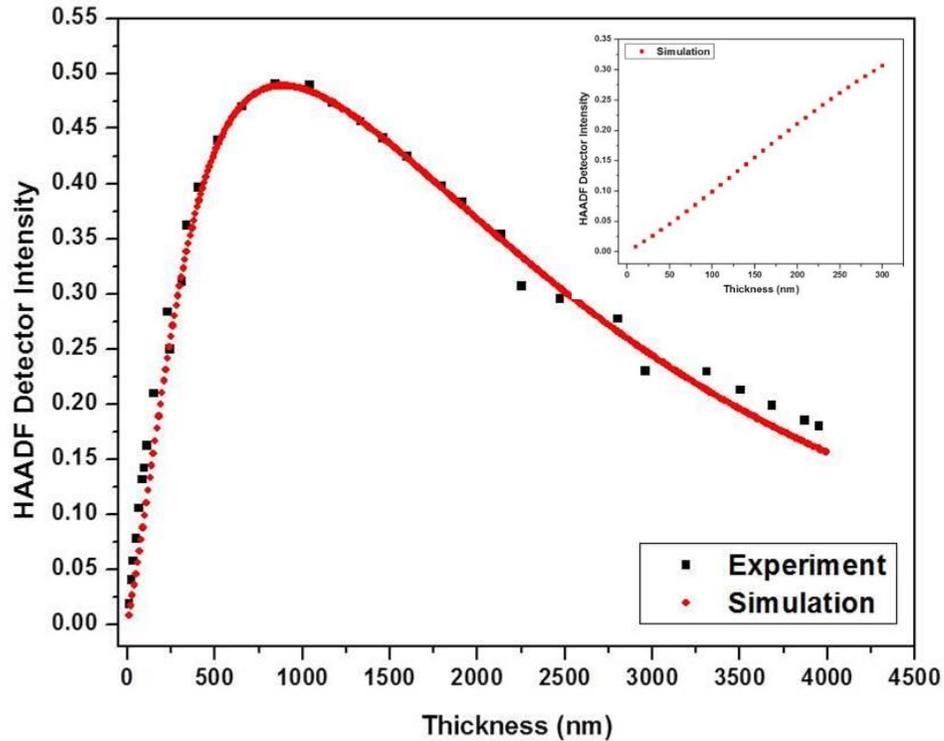

Fig. 7. HAADF-STEM intensity of a wedge shaped Cu sample determined by direct measurments and by simulations. The inset shows the initial slope of simulated curve for the Cu sample using the modified multislice method.

The atomic contrast simulations were performed using the algorithm described in section 3.1. We use the elastic scattering factors for W and SiO2 from *Peng et al.*[18]. The absorption parameter ($\mu$) was determined for W and SiO2 experimentally as described above. A convergent electron beam is defined and for every beam position thousands of electron was made to transverse through the material while the path of the electrons was tracked. A low-density material (SiO2) was used on one side, while a high-density material (W) was used on the other side. For a small number of electrons the simulated paths are shown in Fig. 8. below.

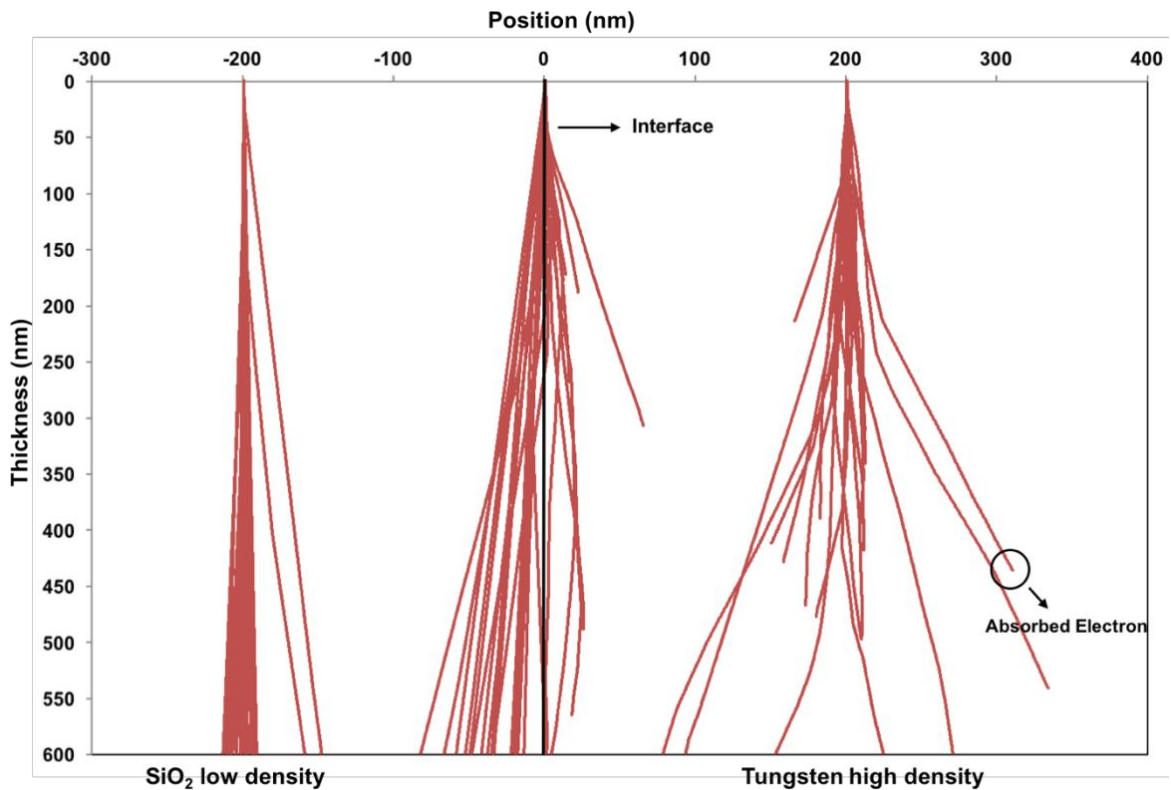

Fig. 8. Simulated electron paths near the interface between two materials

As expected, the electrons get scattered and absorbed more frequently in the high-density material as compared to the low-density material. Near the interface some of the electrons scattered in the high-density material traverse to the lower density material, as shown in Fig. 8. More electrons get absorbed in the higher density material as the electrons lose energy and are not anymore collected by any detector in the diffraction plane. An example of this type of electron has been marked by a red circle in Fig. 8. This program counts the number of electrons scattered and absorbed in the high and low-density material. Electrons scattered at angles between (80-460) mrad are collected by the HAADF-STEM detector, but some electrons are scattered even to higher angles. Electrons which are scattered at even higher angles do not contribute to the detector signal. The results depicted in Fig. 8. were compared with Monte-Carlo simulations from the CASINO software[28]. 1000 electrons electron trajectories were simulated with an acceleration voltage of 300 keV at regions near the interface of a low-density ($SiO_2$) and high-density material (W) and a similar behavior was observed at the interface of the two materials (Appendix A). Fig. 9. shows the magnitude of the scattering angle of some electrons as they pass through the samples. More and more electrons are scattered to higher angles as the thickness of the sample increases.

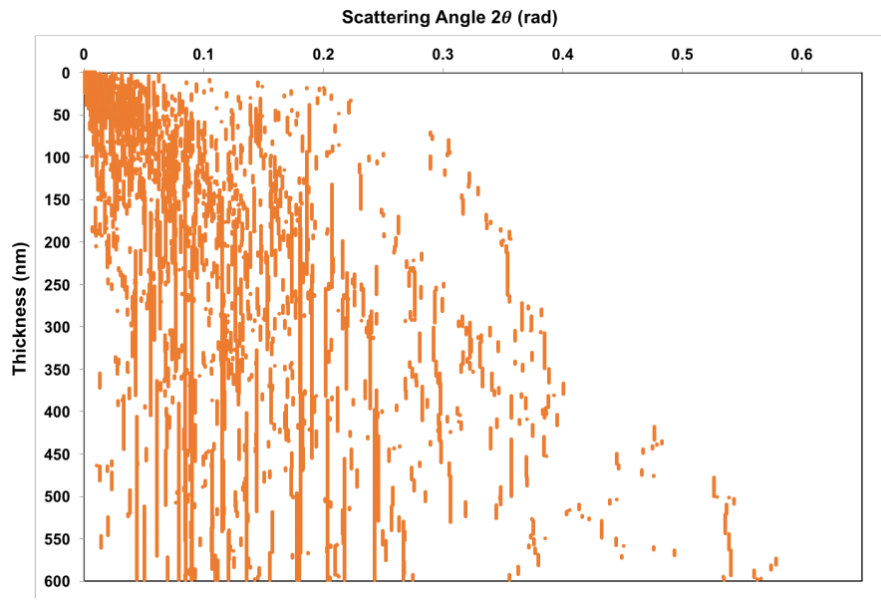

Fig. 9. Magnitude of the scattering angle for individual electrons in a high density material (W) as a function of sample thickness.

For a thick sample the detector intensity is reduced by electrons that are scattered to higher angles than the maximum detection angle of the HAADF-STEM detector (460 mrad).

Fig. 10. (top) shows the variation of the detector-STEM signal with increasing thickness of a sample. For this purpose, we have used a wedge-shaped FIB sample of W and SiO₂ layers. The scattering of electrons from higher-density to lower density material near the interface is not much pronounced at lower thickness but with sample thicknesses greater than 200 nm, this effect is much more noticeable as depicted in Fig. 10. (bottom). Fig. 10. (bottom) also shows that the increase in detector intensity appears mostly within (0-20) nm from the interface of the two materials.

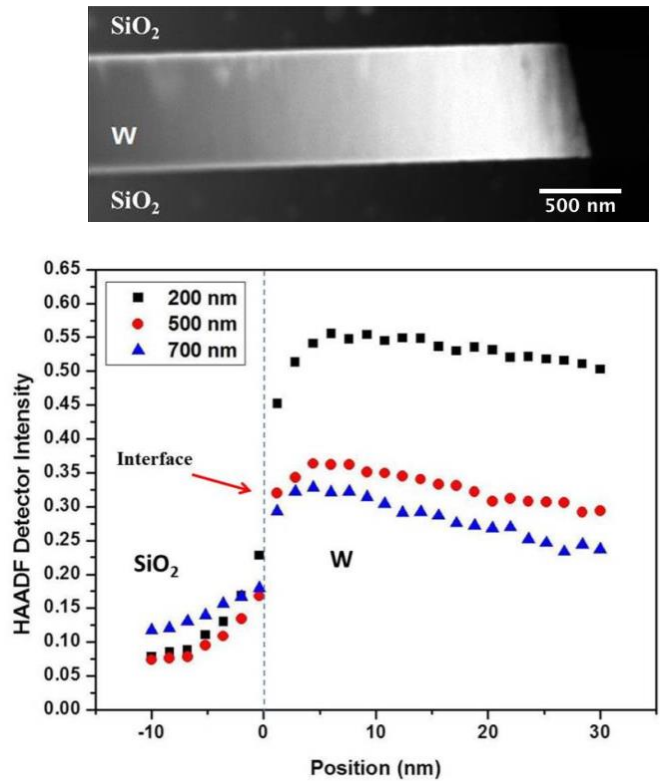

Fig. 10. (Top) HAADF –STEM micrograph showing W and SiO₂ layers for a wedge shaped sample. (Bottom) HAADF – STEM detector signal for varying sample thickness.

Fig. 11. (top) shows a grey scale image generated from the probabilistic model simulations. The black layer represents SiO₂ and the top white layer W. As we move from left to right the simulated

sample thickness increases, which can be seen in the gray scale image and from the profile in Fig. 11. (bottom). The inset in Fig. 11. (bottom) shows the vertical scan across the SiO$_2$-W interface as marked by the red line at 670 nm sample thickness in Fig. 11. (top). There is some statistical noise in the data of this line scan across the simulated interface as 10,000 incident electrons were used for the simulation. Increasing the number of incident electrons in the simulation can reduce statistical variations. Nevertheless, even with this small number of incident electrons used in the simulation, there is an obvious increase in the HAADF-STEM signal near the interface.

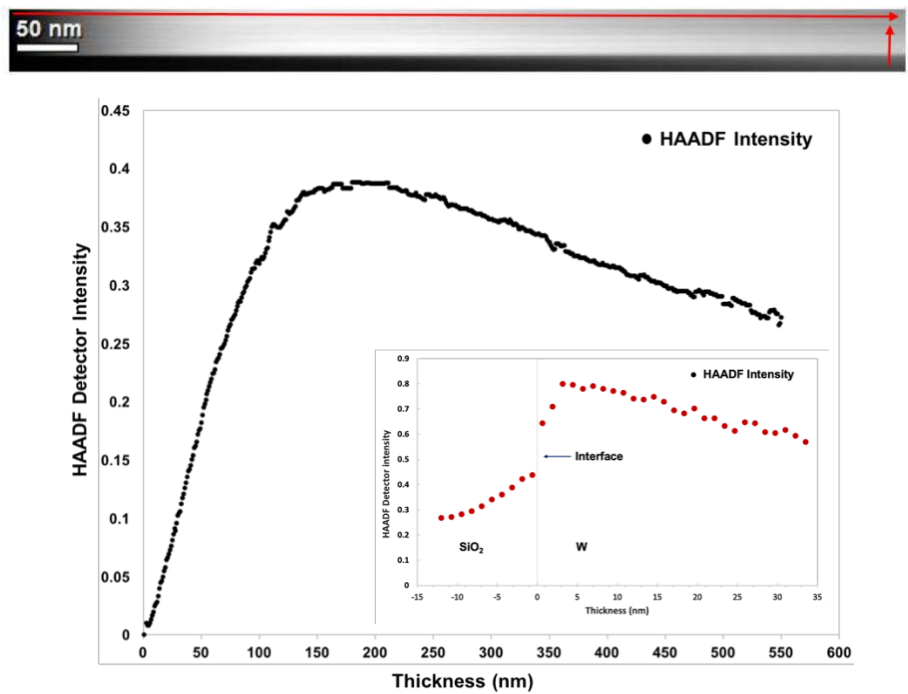

Fig. 11. (Top) Gray scale image formed the probabilistic model simulation. (Bottom) Horizontal line scan for W as marked by the red line. The inset shows the vertical line scan across the SiO$_2$ –W interface as marked by the red line on the top image at a sample thickness of 670 nm.

## 5. Conclusions

This work explains and quantifies the scattering effects occurring in HAADF-STEM measurements near the interface of a high and low-density material in relatively thick TEM samples (100 nm or more). The coefficients for elastic scattering factors computed by *Peng et al.*[18] and a single absorption parameter ($\mu$) (specific to a material) were used to simulate for the atomic number contrast in a probabilistic model where a convergent beam was scanned across the interface of a high and low-density material. The model tracks the path of each of the several thousand electrons transmitted through the material for each beam position.

As the thickness of the sample increases, more electrons are scattered to higher angles resulting in a reduced detector intensity in high-density materials. The simulation results show however, that near the interface of two materials, some of the electrons scattered in the high-density material traverse to the low-density material which accounts for an increased HAADF intensity at the interface in thick samples. Fig. 10. shows that the increase in atomic contrast occurs mostly within 0-20 nm from the interface and is more pronounced for thicknesses greater than 200 nm. These observations were validated both from experimental micrographs (Fig. 10.) and from line scans generated from simulations (Fig. 11.). Thus, our model is capable of explaining the atomic contrast observed at interfaces in relatively thick TEM samples in the HAADF-STEM mode.

**Acknowledgements**

Financial support by Triquint/Qorvo in Apopka, Florida and the Florida High Tech Corridor Industry Matching Research Program is greatly appreciated We acknowledge Qorvo Inc. (Apopka, FL, U.S.A.) for providing us with multilayered samples and financial support. We also thank all the staff members of Advanced Material Processing and Analysis Center at the University of Central Florida (UCF) for providing support and help with handling of the instruments for the measurements. We acknowledge Dr. Bo Yao and Dr. Kevin Coffey from Dept. of Materials Science & Engineering for providing us with multilayered thin film samples.

**Appendix A**: Monte-Carlo simulations using the CASINO SEM software

Thousand electron trajectories were simulated with an electron beam under 300 keV accelerating voltage across two materials of depth 600 nm; SiO2 on the left and W on the right. The atomic densities of SiO2 and W was used from the CASINO database. The convergent beam of electrons was scanned at varying distances from the interface of the two materials as shown in the Fig. 12. In Appendix A. The electron trajectories are colored according to the region across the interface.

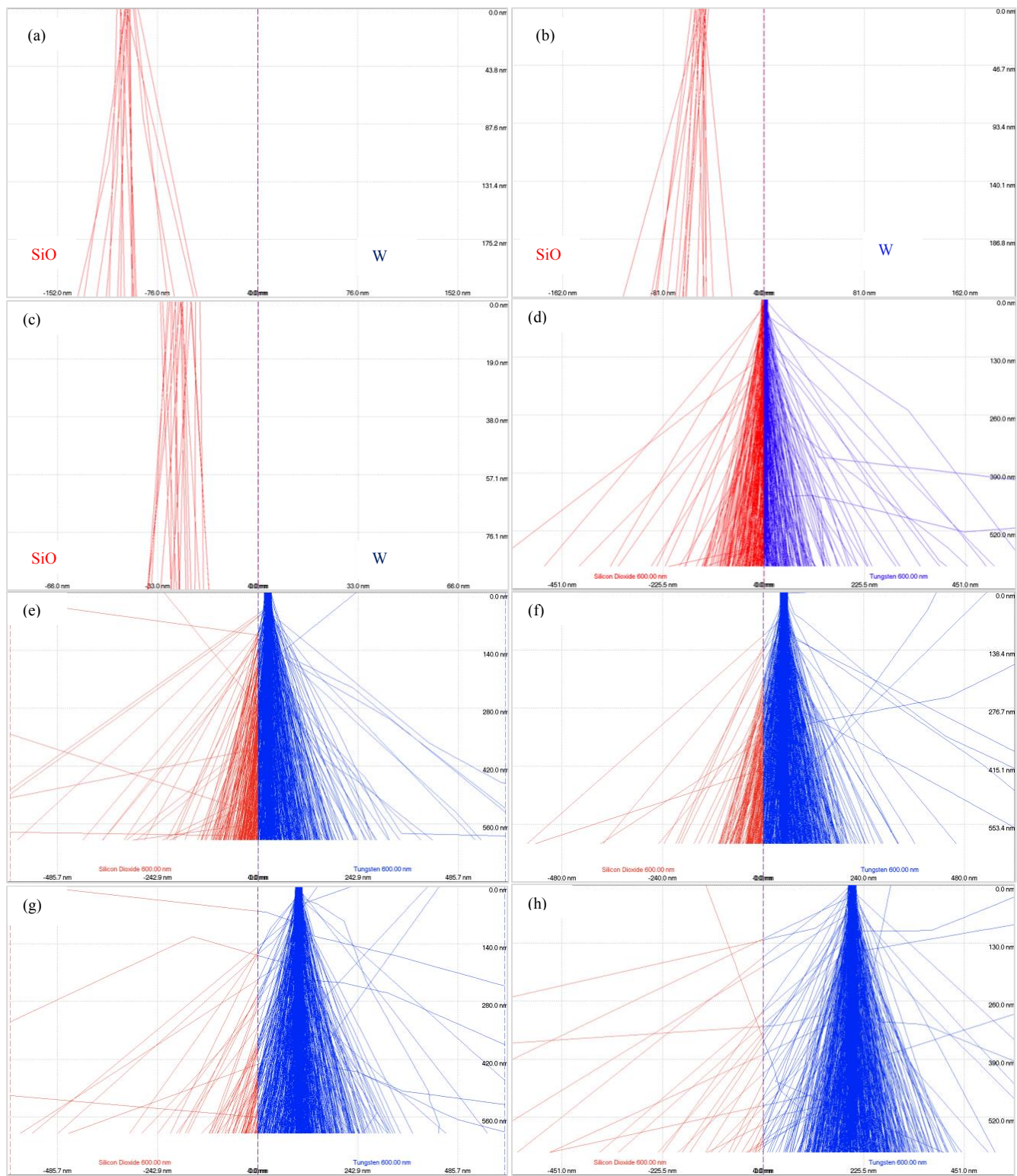

Fig. 12. Simulated electron trajectories at 300 keV from the CASINO software. The dotted lines in the middle represents the interface of two materials, SiO$_2$ (low-density) on the left and W (high-density) on the right. The position of the convergent beam of electrons is at 100 nm (a), 50 nm (b) and 25 nm (c) in SiO$_2$ and at 25 nm (e), 75 nm (f), 100 nm (g) and 200 nm (h) in W. The electron beam is at the interface of the two materials at (d).